\documentclass{article}
\usepackage{spconf,amsmath,epsfig}
\usepackage{amssymb}

\usepackage[numbers]{natbib}

\usepackage[absolute,showboxes]{textpos}

\TPMargin{5pt}

\newcommand{\copyrightstatement}{
	\begin{textblock}{0.84}(0.08,0.93)    
		\noindent
		\scriptsize
		\copyright 2023 IEEE. Published in the IEEE 2023 International Geoscience \& Remote Sensing Symposium (IGARSS 2023), scheduled for July 16 - 21, 2023 in Pasadena, CA, USA. Personal use of this material is permitted. However, permission to reprint/republish this material for advertising or promotional purposes or for creating new collective works for resale or redistribution to servers or lists, or to reuse any copyrighted component of this work in other works, must be obtained from the IEEE. Contact: Manager, Copyrights and Permissions / IEEE Service Center / 445 Hoes Lane / P.O. Box 1331 / Piscataway, NJ 08855-1331, USA. Telephone: + Intl. 908-562-3966.
	\end{textblock}
}

\title{DEALING WITH NON-GAUSSIANITY OF SAR-DERIVED WET SURFACE RATIO FOR FLOOD EXTENT REPRESENTATION IMPROVEMENT}
\name{T.~H.~Nguyen$^{1,2}$, S.~Ricci$^{1,2}$, A.~Piacentini$^{1}$, E.~Simon$^{3}$, R.~Rodriquez~Suquet$^{4}$ and S.~Peña~Luque$^{4}$
\thanks{Thanks to Space for Climate Observatory for funding.}}

\address{\small $^{1}$Centre Européen de Recherche et de Formation Avancée en Calcul Scientifique (CERFACS), 31057 Toulouse Cedex 1, France\\
    \small $^{2}$CNRS, CECI/CERFACS, UMR 5318, 31057 Toulouse Cedex 1, France\\
	\small $^{3}$ENSEEIHT, Toulouse INP, 31071 Toulouse Cedex 7, France\\
    \small $^{4}$Centre National d'Études Spatiales (CNES), 31401 Toulouse Cedex 9, France
 }

\begin{document}
\copyrightstatement
%
\maketitle
\begin{abstract}
Owing to advances in data assimilation, notably Ensemble Kalman Filter (EnKF), flood simulation and forecast capabilities have greatly improved in recent years. The motivation of the research work is to reduce comprehensively the uncertainties in the model parameters, forcing and hydraulic state, and consequently improve the overall flood reanalysis and forecast capability, especially in the floodplain. It aims at assimilating SAR-derived (typically from Sentinel-1 mission) flood extent observations, expressed in terms of wet surface ratio.
The non-Gaussianity of the observation errors associated with the SAR flood observations violates a major hypothesis regarding the EnKF and jeopardizes the optimality of the filter analysis. Therefore, a special treatment of such non-Gaussianity with a Gaussian anamorphosis process is thus proposed. This strategy was validated and applied over the Garonne Marmandaise catchment (South-west of France) represented with the TELEMAC-2D hydrodynamic model, focusing on a major flood event that occurred in December 2019. The assimilation of the SAR-derived wet surface ratio observations, in complement to the in-situ water surface elevations, is illustrated to consequentially improve the flood representation.
\end{abstract}

\section{Introduction}
\label{sec:intro}
Floods cause widespread devastation and result in losses of life and damages to properties and critical public security infrastructures. Between 1998-2017, floods directly expose nearly 20\% of the world population \cite{rentschler2020people}.
2D hydrodynamic numerical codes have been developed to predict water surface elevation (WSE) and velocity for a lead time varying from a few hours to several days, thereby allowing to assess flood risks effectively. However, these models are still imperfect due to  uncertainties within the models and the inputs, e.g., friction coefficients and boundary conditions.
Data assimilation (DA) aims at estimating the optimal state of a model by sequentially combining the model and the observations while taking into account their respective uncertainties \cite{asch2016data}. Owing to advances of DA in hydrology and hydrodynamic, notably Ensemble Kalman Filter (EnKF), flood simulation and forecast capabilities have greatly improved in recent years. Nonetheless, the effectiveness of DA strongly depends on the characteristics of the observing network, namely the spatial and temporal density of the observations, as well as the statistics of observation errors. Given the increasing volume and  diversity of remote-sensing observations of continental waters, there is a great interest for an effective and harmonizing assimilation of heterogeneous data from synthetic aperture radar (SAR), optical and altimetry satellite missions. In this paper, we present a pre-processing step to treat non-Gaussian SAR-derived observations with a Gaussian anamorphosis (GA) applied to the EnKF analysis. 
It allows mapping the control and/or observation variables onto a transformed space where the Gaussianity assumption is better satisfied.

\section{State of the Art}
\label{sec:related}
\subsection{Remote-sensing flood observations}
Satellite SAR data is particularly advantageous as it allows an all-weather day-and-night global-coverage imagery of continental water. They are identified by low-backscatter pixels resulted from the specular reflection of the incident radar pulses \cite{martinis2015flood}. 
Most commonly used strategies are similar to \cite{giustarini2011assimilating}, that is from SAR images taken during flood events, flood edges are identified and  integrated with an available Digital Elevation Model (DEM) to derive the WSE within the floodplain. They are then assimilated to the WSE simulated by 1D/2D hydrodynamic model to sequentially update the model states and parameters. 

Many existing strategies, such as \cite{revilla2016integrating} and \cite{lai2014variational}, depend on the expression of flood extents as a function of the hydraulic state, whereas other research works \cite{hostache2018near,cooper2019observation} propose a more direct use of SAR observations. For instance, \cite{hostache2018near} presents the assimilation of SAR-derived flood probability maps using a Particle Filter (PF) with a sequential importance sampling built on a coupled hydrologic-hydraulic model. 
Such a map represents the probability of an observed backscatter value being flooded, assuming that its prior probability to be flooded or non-flooded are two Gaussian probability density functions \cite{giustarini2016probabilistic}. On the other hand, 
\cite{cooper2019observation} proposed an observation operator to treat synthetical SAR backscatter values directly as the observations to circumvent the processes of flood mapping and/or flood probability estimation.

\subsection{Dealing with non-Gaussianity in DA}
The optimality of the Kalman filters (KF) and variational analysis relies on the Gaussian assumptions for the background and observation errors, as well as on the linearity of the observation operator that relates the control and the observation spaces \cite{asch2016data}. When these assumptions are not fulfilled, the KF or variational analysis can still be done but they are suboptimal. Therefore, the non-Gaussianity characteritics of SAR-derived observation errors need to be properly accounted for in the framework of DA. 

The non-Gaussianity of the control and/or observation errors can be handled using a DA algorithm that does not require Gaussianity assumptions. For instance, PF works with the entire probability function, instead of focusing on the first and second moments of the statistics like KF and variational algorithm. Such a solution has been considered by a number of studies, favoring a PF or a Bayesian approach to assimilate SAR-derived observations.
Indeed, the PF framework used in \cite{giustarini2011assimilating,hostache2018near} offers the key advantage of relaxing the assumption that observation errors are Gaussian, and allows to propagating a non-Gaussian distribution through non-linear hydrologic and hydrodynamic models.  
This makes it better suited for a DA of probabilistic flood maps than the more widely used EnKF \cite{revilla2016integrating} or variational approaches \cite{lai2014variational}.
Otherwise, when the Gaussianity assumptions are violated, a pre-processing step is necessary.

\section{Method}
\label{sec:method}
\begin{figure}[t]
    \centering\includegraphics[width=0.5\textwidth]{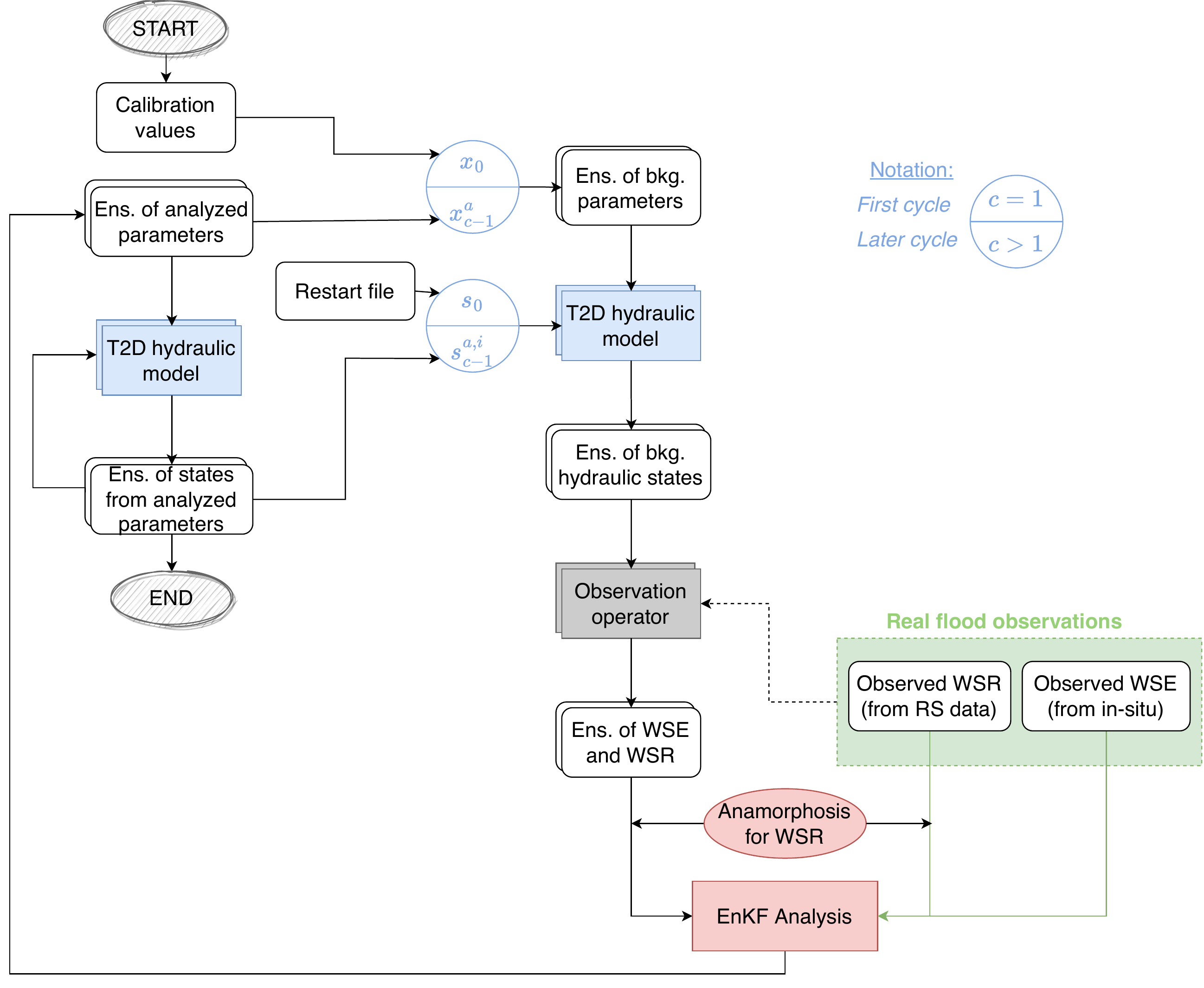}
    \caption{Workflow of the GA-EnKF DA strategy.}
    \label{fig:ExpSetWorkflow}
\end{figure}

Fig. \ref{fig:ExpSetWorkflow} presents the main steps of the proposed strategy. Here, an EnKF with a dual state-parameter analysis is implemented on top of a TELEMAC-2D (T2D) hydrodynamic model for the Garonne Marmandaise river reach, similar to our previous work \cite{nguyenagu2022}. The EnKF control vector is composed of spatially-distributed friction coefficients and a corrective parameter of the inflow discharge; it is augmented with a uniform correction of the hydraulic state within the floodplain subdomains. This DA strategy was performed, validated and assessed over the major flood event that occurred in December 2019.

This research work focuses on the assimilation of flood extent observations derived from Sentinel-1 SAR images. Using SAR images, a flood extent mapping method is achieved by applying a Random Forest algorithm which was trained on past flood events from the Copernicus Emergency Management Service Rapid Mapping directory for which flood maps were manually delineated \cite{kettig}. The resulting binary wet/dry water masks are further expressed in terms of wet surface ratios (WSR) computed over chosen subdomains of the floodplain. Such a ratio, despite being non-Gaussian and bounded within $[0,1]$, is assimilated jointly with in-situ WSE gauge observations, shown by green box in Fig. \ref{fig:ExpSetWorkflow}, to improve the flow dynamics within the floodplain, which yielded promising results \cite{nguyenagu2022}. 

Full details on the performed cycled EnKF can be found in \cite{nguyenagu2022,NguyenTGRS2022}. 
In addition, the proposed  Gaussian Anamorphosis (GA) technique, also known as normal-score transform, stands in transforming the distribution of the SAR-derived observations into a Gaussian distribution, compatible with KF-based DA algorithms. This alleviates the need for advanced processes on the PF implementation to avoid ensemble collapse.  GA was proposed by \cite{bertino2003} and it has been since investigated in different works.
In the present work, it is applied to the model parameter analysis, instead of the state analysis as in most studies.
GA involves transforming the state variables and observations into new variables with Gaussian features, over which the DA analysis is computed. 

\begin{figure}[!h]
    \centering
    \includegraphics[width=0.8\linewidth]{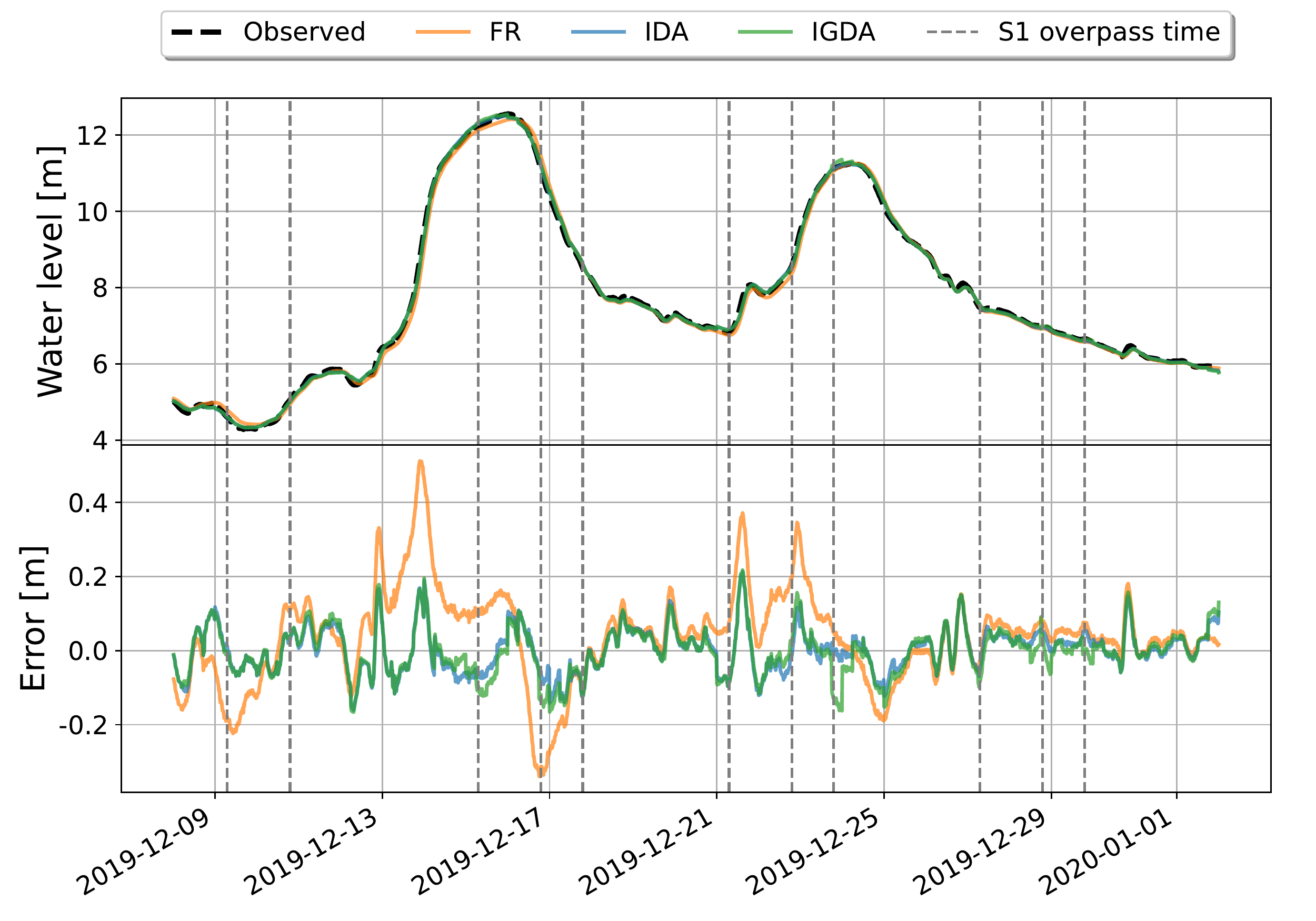}

	\includegraphics[width=0.8\linewidth]{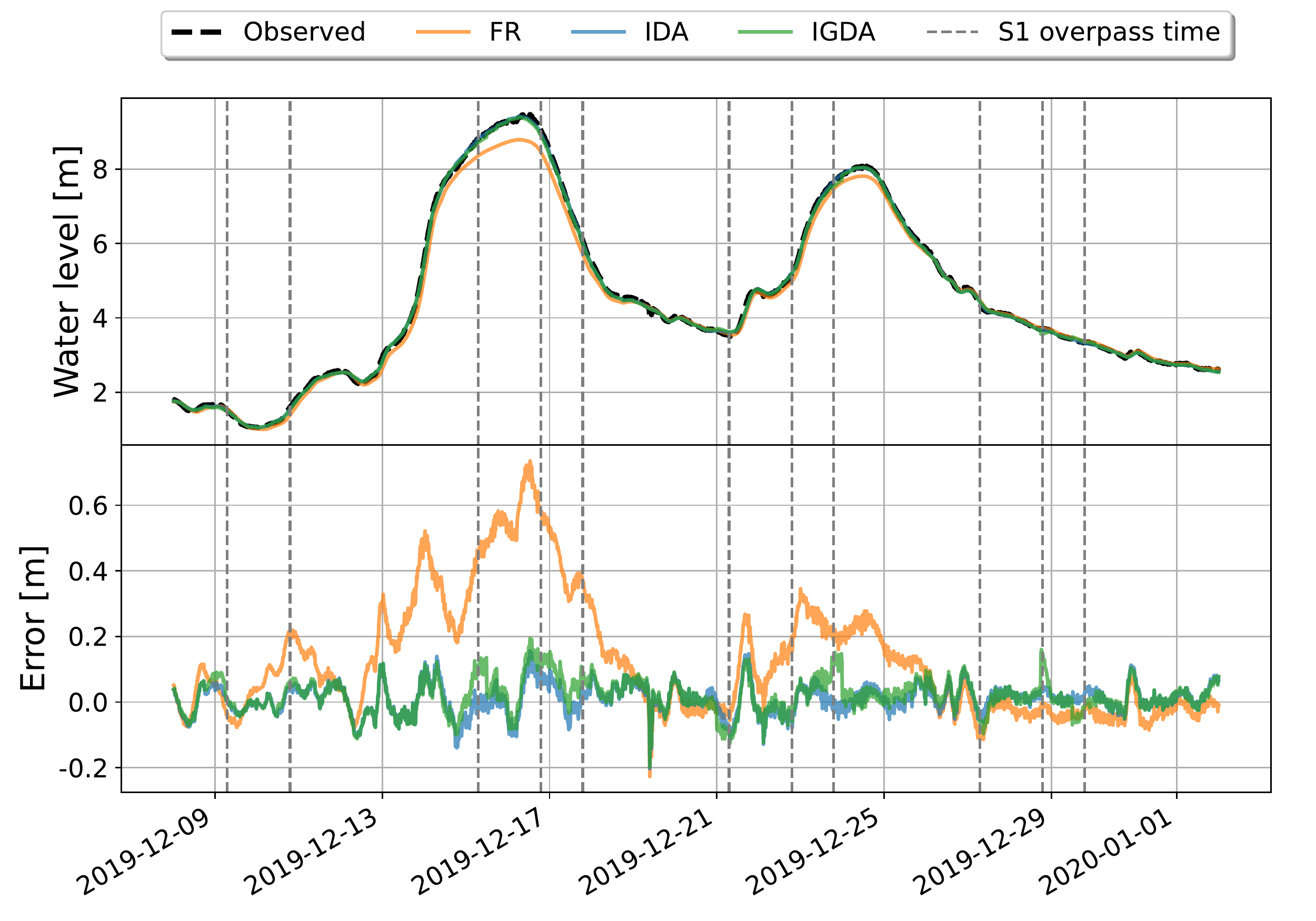}
	
	\includegraphics[width=0.8\linewidth]{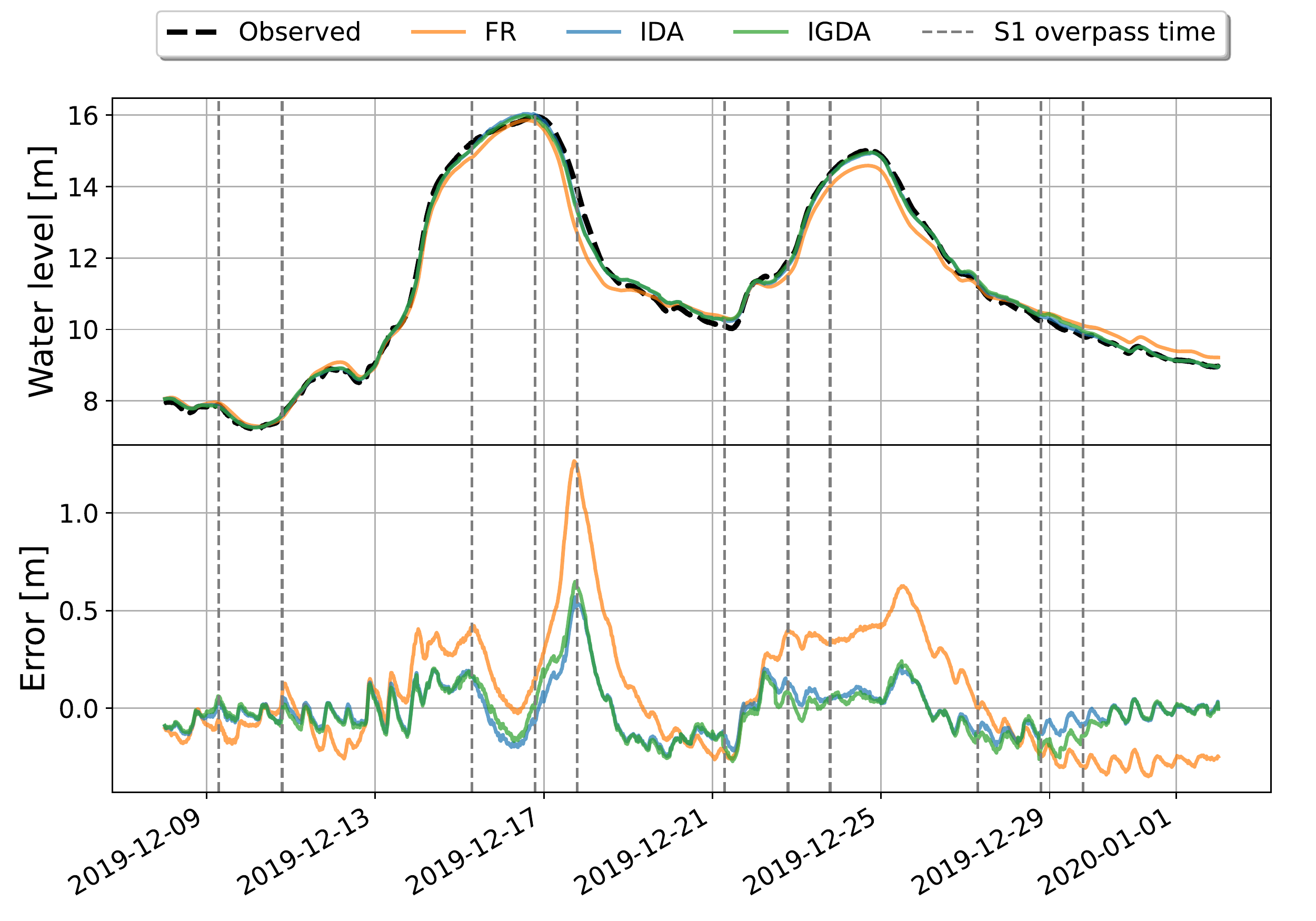}
    \caption{Water levels at in-situ observing stations (a) Tonneins, (b) Marmande, (c) La Réole. Top panel: simulated WSE vs. observed WSE; bottom panel: Observation-minus-Analysis errors.}\label{fig:1D-assessment}
\end{figure}

\begin{figure}[!h]
    \includegraphics[width=0.8\linewidth]{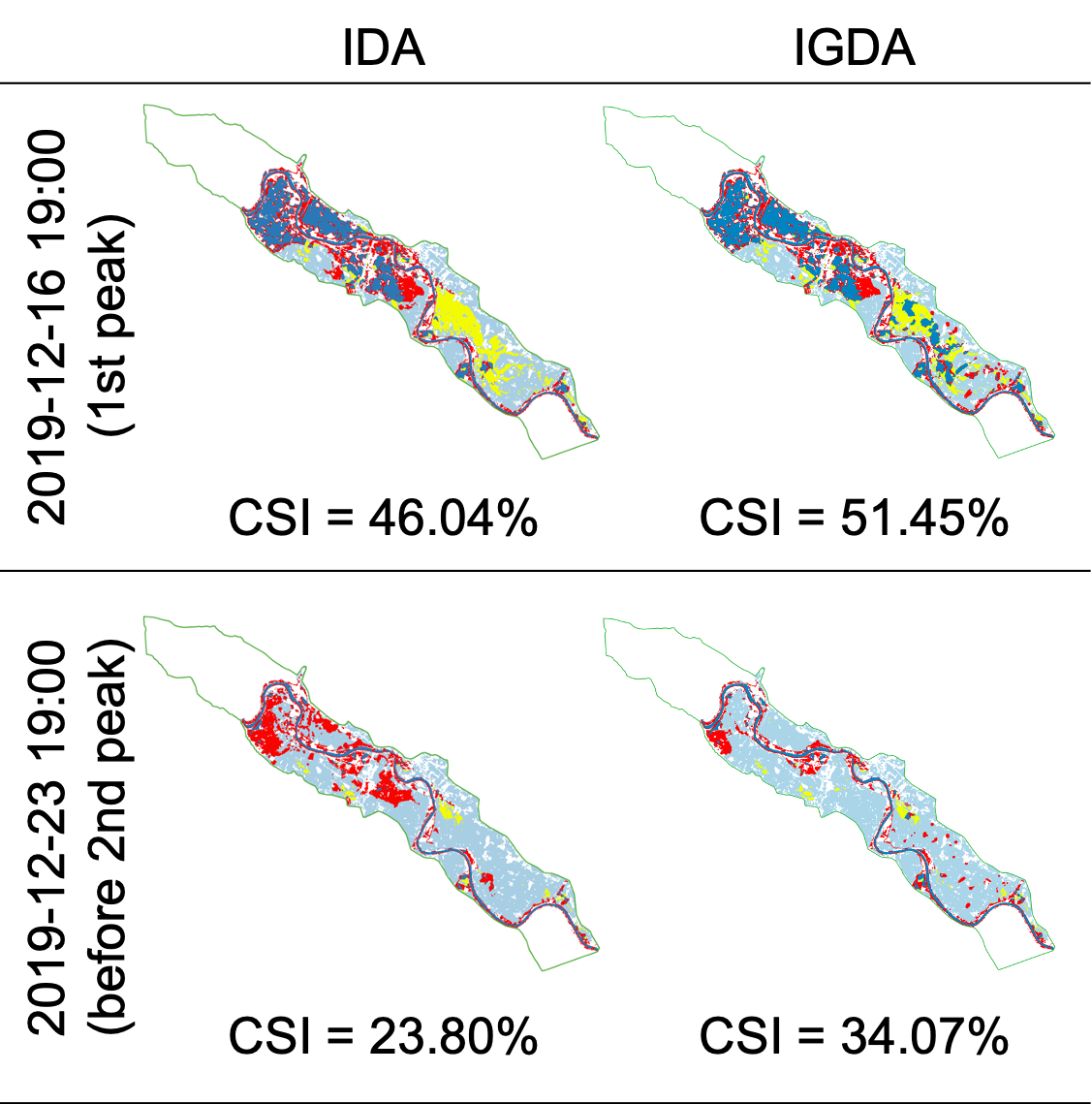}

    \includegraphics[width=0.6\linewidth]{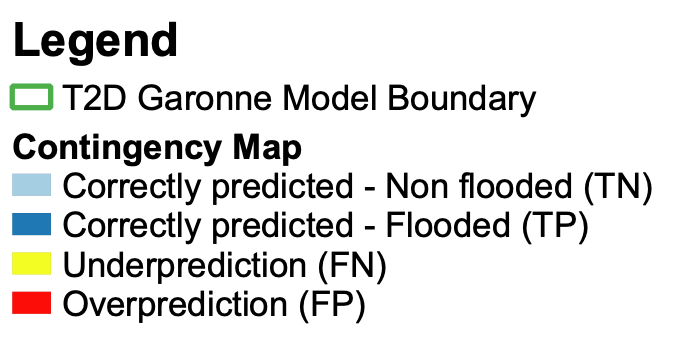}
    \caption{Contingency maps between simulated flood extent and observed flood extent derived from S1 image.}\label{fig:2D-assessment}
\end{figure}

The anamorphosis function $\phi_c$ for a particular cycle $c$ is devised from the empirical marginal distribution of the variable in the observation space. For such a purpose, all observations in time and space over the respective window are taken into account altogether. The three-step algorithm to set up the anamorphosis function is described in \cite[Appendix A]{simon2009}. The DA analysis can thus be carried out in the transformed space.
For cycle $c$, the anamorphosis function $\phi_c$ is a non-linear bijective function, that maps WSR physical space to a Gaussian space:
\begin{equation}
    \tilde{{\bf y}}^{o,i}_{c, \mathrm{WSR}} = \phi_c \left({\bf y}^{o,i}_{c, \mathrm{WSR}} \right), \; \tilde{{\bf y}}^{f,i}_{c, \mathrm{WSR}} = \phi_c \left({\bf y}^{f,i}_{c, \mathrm{WSR}} \right).
\end{equation}

This comes down to redefining the observation operator ${\cal{H}}_c$ as $\tilde{{\cal{H}}}_c$ that now maps the hydraulic state variable ${\mathbf{s}}^{f,i}_{c}$ onto the transformed space thanks to the GA:
\begin{equation}
{\tilde{\mathbf{y}}}^{f,i}_{c} = {\tilde{{\cal{H}}}_c}\left({\mathbf{s}}^{f,i}_{c}\right) = \phi_c \left({{\cal{H}}}_c\left({\mathbf{s}}^{f,i}_{c}\right) \right)
\end{equation}
where ${{\tilde{\cal{H}}}_c}$: ${\mathbb{R}}^m \rightarrow {\mathbb{R}}^{n_{obs}}$ selects, extracts and interpolates model outputs at times and locations of the in-situ WSE observations $\mathbf{y}^o_{c, \mathrm{H}}$, whereas it selects, extracts and applies the anamorphosis function $\phi_c$ at times and locations of WSR observations ${\bf y}^{o,i}_{c,\mathrm{WSR}}$, over $W_c$. This corresponds to prescribing an identity function for the anamorphosis of the in-situ WSE observations and $\phi_c$ for that of the WSR observations.

The assimilation of SAR-derived WSR observations, in complement to in-situ observations, is carried out with the devised GA-assisted EnKF.
The EnKF analysis step (red ellipse in Fig. \ref{fig:ExpSetWorkflow}) stands in the update of the control ${\bf x}_c^{a,i}$ and the associated model state vector ${\bf s}_c^{a,i}$, here achieved in the anamorphosed space and represented by a red rectangle in Figure~\ref{fig:ExpSetWorkflow}. This differs from the classical EnKF analysis performed in previous works \cite{nguyenagu2022,NguyenTGRS2022}, as the computation of the innovations and the covariance matrices are performed in the transformed space using the transformed observation operator ${{\tilde{\cal{H}}}_c}$:
\begin{equation}
{\bf x}_c^{a,i} = {\bf x}^{f,i}_{c} + \mathbf{K}_{c} \left(\tilde{\bf y}^{o,i}_{c} - \tilde{\bf y}^{f,i}_{c}\right).
\label{eq:ctlana}
\end{equation}

The Kalman gain is further computed from covariance matrices stochastically estimated within the ensemble, considering anamorphosed observation vectors  $\tilde{\bf y}^{f,i}$  instead of ${\bf y}^{f,i}_{c}$.
\begin{equation}
	\mathbf{K}_c = \mathbf{P}^{\bf{x},\tilde{\bf{y}}}_c {\left[ \mathbf{P}^{\tilde{\bf{y}},\tilde{\bf{y}}}_c + \mathbf{R}_{c} \right]}^{-1}.
	\label{eq:EnKF_ana_Klambda_gain_chap12}
\end{equation}

The proposed method has been validated in Observing System Simulation Experiment and with a major flood event in January-February 2021, which can be found in \cite{nguyen2023gaussianarxiv}.

\section{Results}
\label{sec:results}
This paper presents the merits of assimilating 2D flood observations derived from Sentinel-1 SAR images with an EnKF implemented
on the T2D hydrodynamic model, for a flood event in 2019. The observation is expressed in terms of Wet Surface Ratio or WSR computed over defined sensitive floodplain subdomains. These non-Gaussian measurements are processed with a Gaussian anamorphosis.
Three experiments are carried out: a free run without assimilation (FR), and two DA experiments, namely IDA that assimilates only in-situ WSE measurements and IGDA that assimilates both in-situ WSE and remote-sensing WSR observations. 

Table \ref{tab:RMSE} summarizes the quantitative 1D and 2D metrics computed to assess the merits of the GA-EnKF strategy on the estimation of WSE time-series at gauge stations in the river bed and the representation of the flood extents in the floodplain. Fig. \ref{fig:1D-assessment} depicts the 1D assessments between simulated WSE and observed WSE at three observing stations. The top panel displays the observed WSE (black-dashed line) and the simulated water level time-series simulated by FR (orange curve) and by the DA experiments (IDA in blue and IGDA in green). The bottom panel displays the discrepancies between the observations and the simulated WSE. The RMSEs of WSE at observing stations show a significant reduction achieved by the EnKF, with and without the GA.

The contingency maps and Critical Success Index (CSI) computed with respect to the SAR-derived water mask are shown in Fig. \ref{fig:2D-assessment} for both IDA and IGDA. It should be noted that the contingency maps for FR are quite similar to those of IDA. While the CSI (Table \ref{tab:RMSE}) is slightly decreased from FR to IDA, it is significantly improved by the IGDA with respect to FR and IDA. Both 1D and 2D analysis show that the proposed DA strategy significantly improves the flood dynamics with respect to the FR, especially during the flood recess. It also shows that the treatment of non-Gaussianity brings further improvements. 
It is worth-noting that such effectiveness regarding the flood recess is beneficial for this event with a complex dual-peak dynamic.
This work pushes for a reliable strategy for flood forecasting over poorly-gauged catchments that relies on a proper assimilation of remote-sensing-derived flood extent observations in spite of the non-Gaussianity of their observation errors.

\begin{table}[t]
    \centering
    \small
    \caption{1D RMSE with respect to in-situ WSE measured at observing stations, and 2D CSI with respect to S1-derived flood extent maps.}\label{tab:RMSE}
    \begin{tabular}{c|ccc|cc}
        \hline
        & \multicolumn{3}{c|}{{Root-mean-square error [m]}} & \multicolumn{2}{c}{{CSI [\%]}}\\
        \hline
        Exp. & Tonneins & Marmande & La Réole & 16 Dec & 23 Dec \\
        \hline
        {FR} & 0.129 & 0.220 & 0.318 & 46.49 & 24.48 \\
        \hline
        {IDA} & 0.060 & 0.045 & 0.125 & 46.04 & 23.80 \\
        \hline
        {IGDA} & 0.065 & 0.055 & 0.140  & 51.45 & 34.07 \\
        \hline
    \end{tabular}
\end{table}

{\footnotesize
\bibliographystyle{ieeetr}
\bibliography{refs.bib}}

\end{document}